\documentclass[apj]{emulateapj}
\usepackage{cprotect}
\usepackage{footnote}
\bibpunct{(}{)}{;}{a}{}{,}

\begin{document}
% Title Page
\title{Investigating the differential emission measure and energetics of microflares with combined SDO/AIA and RHESSI observations}

\author{A. R. Inglis\altaffilmark{1}, S. Christe}
\affil{Solar Physics Laboratory, Heliophysics Science Division, NASA Goddard Space Flight Center, Greenbelt, MD, 20771, USA}

\altaffiltext{1}{Physics Department, The Catholic University of America, Washington, DC, 20664, USA}

\begin{abstract}
An important question in solar physics is whether solar microflares, the smallest currently observable flare events in X-rays, possess the same energetic properties as large flares. Recent surveys have suggested that microflares may be less efficient particle accelerators than large flares, and hence contribute less nonthermal energy, which may have implications for coronal heating mechanisms. We therefore explore the energetic properties of microflares by combining Extreme Ultraviolet (EUV) and X-ray measurements.

We present forward-fitting differential emission measure (DEM) analysis of 10 microflares. The fitting is constrained by combining, for the first time, high temperature RHESSI observations and flux data from SDO/AIA. Two fitting models are tested for the DEM; a Gaussian distribution and a uniform DEM profile. A Gaussian fit proved unable to explain the observations for any of the studied microflares. However, 8 of 10 events studied were reasonably fit by a uniform DEM profile. Hence microflare plasma can be considered to be significantly multi-thermal, and may not be significantly peaked or contain resolvable fine structure, within the uncertainties of the observational instruments. 

The thermal and non-thermal energy is estimated for each microflare, comparing the energy budget with an isothermal plasma assumption. From the multithermal fits the minimum non-thermal energy content was found to average approximately 30\% of the estimated thermal energy. By comparison, under an isothermal model the non-thermal and thermal energy estimates were generally comparable. Hence, multi-thermal plasma is an important consideration for solar microflares that substantially alters their thermal and non-thermal energy content.

\end{abstract}

\keywords{Sun: corona - Sun: flares}
\maketitle

\section{Introduction}
\label{intro}

The size distribution of solar flares is well known to obey a power-law form \citep[e.g.][]{1971SoPh...16..152D, 2012ApJ...754..112A}; thus large powerful events occur much less frequently than small energy releases. However, despite this relationship, it is unclear whether small events exhibit energetic properties consistent with large flares. Solar microflares \citep[see][for a review]{2011SSRv..159..263H}, generally defined as events of ~ GOES-class C, are the smallest currently observable events in the X-ray regime, and allow us to address this question. Recent surveys of microflares have suggested that, in many respects, microflares are similar to larger flares \citep{Christe2008micro, hannah2008, 2011SSRv..159..263H}. They occur in active regions and their frequency distribution is similar to large flares. Yet in one important aspect they differ significantly; their spectra above 10 keV (usually interpreted as non-thermal emission) are generally steep compared to large flares; interpreted as a power-law the spectral index of microflares is generally between $-5$ and $-8$ \citep[see][]{Benz2002micro, Krucker2002, Christe2008micro, hannah2008} while for large flares it typically ranges from $-2.5$ to $-4$, \citep{Saint-Hilaire2008}, although there are some notable exceptions \citep{2008A&A...481L..45H, 2013A&A...555A..21O}. A recent full review of microflare properties may be found in \citet{2011SSRv..159..263H}. One possible interpretation of this finding is that microflares are not as efficient particle accelerators as large flares. An important consequence of the steepness of microflare spectra is that the total energy in 
non-thermal electrons is strongly dependent on the lower energy cutoff of the non-thermal spectra which is poorly constrained due to thermal emission at low energies. 

Historically the thermal emission from microflares has been interpreted as isothermal despite the fact that it has been shown (e.g., \citealt{mctiernan1999}) that microflare plasma may be multi-thermal. Modeling a multi-thermal plasma can provide a more accurate estimate of the thermal energy content. Additionally it also provides a more accurate measure of the low energy cutoff and therefore of the non-thermal energy content. These two facts combined suggest that better insight into microflare energetics and therefore the energy input into the solar corona may be gained by considering multithermal models of the microflare plasma.

In this work, we analyze the energy content of ten microflares by combining observations in X-rays and EUV from SDO/AIA and RHESSI. Through these combined observations the thermal emission is modeled with a multi-thermal differential emission measure instead of an isothermal model. We present an analysis of the energy content of these microflares and compare the estimated energy budgets with those obtained under isothermal assumptions.

\section{Instruments and data selection}

\begin{figure}[h]
 \begin{center}
  \includegraphics[width=8.5cm]{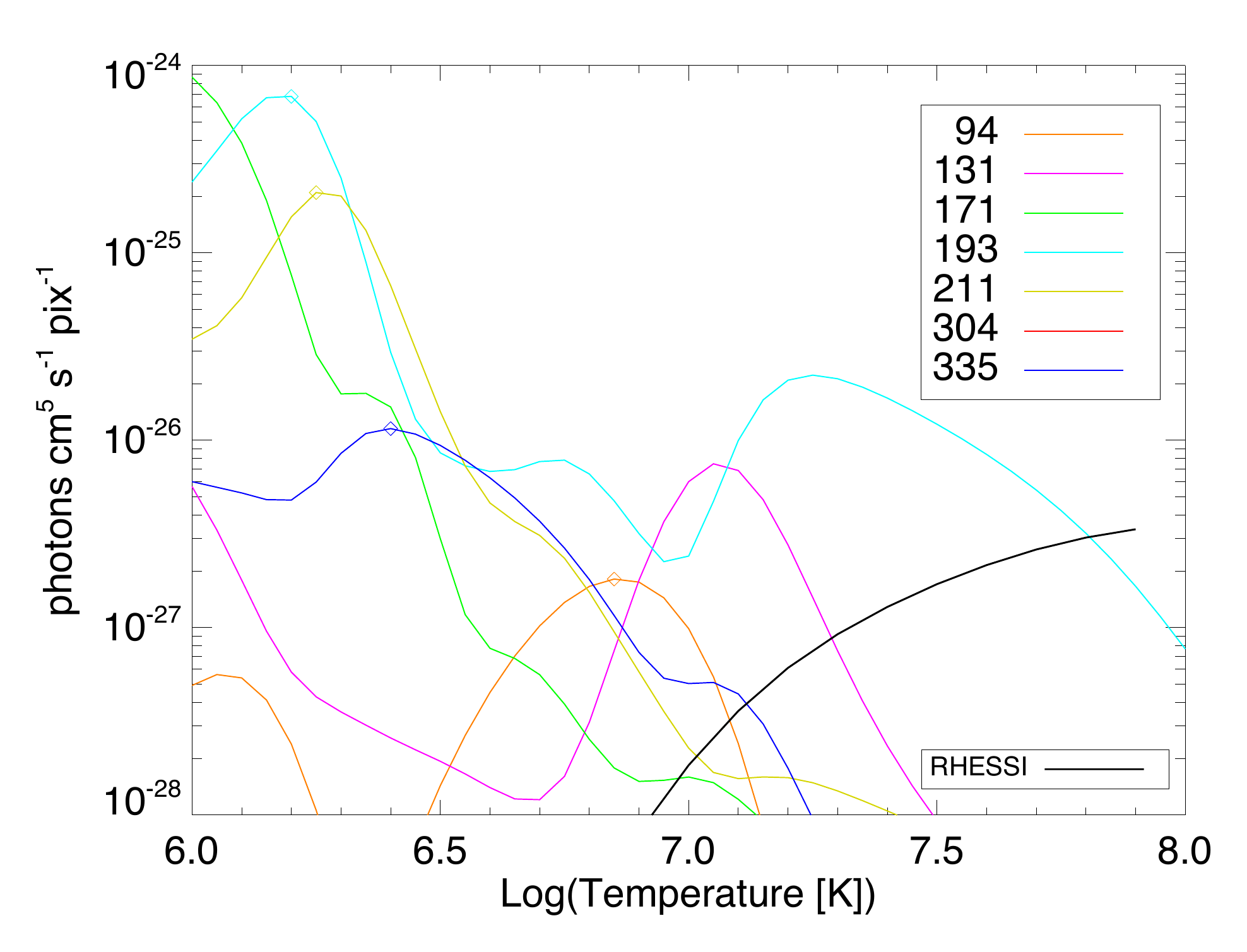}
  \cprotect\caption{Temperature response functions for the seven EUV channels of AIA (coloured lines), and an equivalent temperature response function for RHESSI (black line). The RHESSI curve is scaled to one AIA pixel. The AIA temperature response functions include the empirical \verb|chiantifix| correction, and are obtained using the SolarSoftWare (SSW) procedure \verb|aia_get_response.pro|.}
  \label{response_curves}
 \end{center}
\end{figure}

Launched in February 2010, the Solar Dynamics Observatory \citep[SDO;][]{2012SoPh..275....3P} has provided a new perspective on solar activity. The Atmospheric Imaging Assembly \citep[AIA;][]{2012SoPh..275...17L} on board SDO observes the full Sun at seven extreme ultraviolet (EUV) wavelengths; 94\AA, 131\AA, 171\AA, 193\AA, 211\AA, 304\AA, and 335\AA, covering a temperature range from $<$1 to $>$10 MK. For each of these EUV wavelengths, AIA takes images with a pixel size of 0.6 arcsec at a cadence of 12 s, an unprecedented combination of spatial and temporal resolution for full-disk solar imaging. Each of AIA's observing wavelengths is associated with a featured, multi-thermal temperature response function. The 171\AA\ channel for example, has peak sensitivity to plasma with a temperature of approximately $T$= 0.6 MK, whereas the 94\AA\ channel is sensitive to plasma at both $T$=1 MK and $T$= 6 MK, and the 335\AA\ channel is sensitive over a broad range from cool plasma as low as $T$= 100, 000 K to hot plasma of $T$= 5 MK. The response functions are shown in Figure \ref{response_curves}.

The most sensitive solar X-ray observations currently available are provided by the Reuven Ramaty High Energy Solar Spectroscopic Imager \citep[RHESSI; ][]{Lin2002}. Launched in 2002, RHESSI has provided high spatial and energy resolution observations of solar flares for over a decade, observing from 3 keV - 17 MeV. RHESSI has sub-second temporal resolution, and produces images of X-ray emission via rotating collimators, with a spatial resolution down to 2.3 arcsec.

RHESSI provides excellent complementary data for temperature analysis of high temperature solar plasma. As Figure \ref{response_curves} shows, with the exception of 193\AA\, the AIA channels are not very sensitive to plasma beyond $T$= 10 - 20 MK compared to lower temperatures, while RHESSI's lower sensitivity limit is $T$ $\approx$ 8 MK, and extends far higher than AIA (Figure \ref{response_curves}, black line). Thus RHESSI is ideal for constraining the properties of high temperature plasma. In the energy range of interest for microflares (3 - 30 keV) and active regions, the energy resolution of RHESSI spectral data is around 0.33 keV.

In this paper, we focus on the smallest currently observable events with characterizable X-ray emission, so-called microflares. X-rays provide important information about the high temperature plasma and provide direct observations of accelerated electrons present in microflares, allowing us to better constrain their energy content. For this work we analyze a set of ten microflares randomly chosen from the RHESSI flare catalogue. All of these events were observed by both SDO/AIA and RHESSI between 2011 and 2012. To avoid saturation effects in AIA all of these events are of GOES-class B or lower. For reference throughout this paper these events are numbered sequentially according to their date, with $\#1$ representing the earliest event. Details of the event dates, times, and the chosen fit intervals are listed in Table \ref{fit_table}.

\section{Estimating the differential emission measure (DEM)}
\label{fitting_dems}

\begin{figure*}
\begin{center}
\includegraphics[width=14cm]{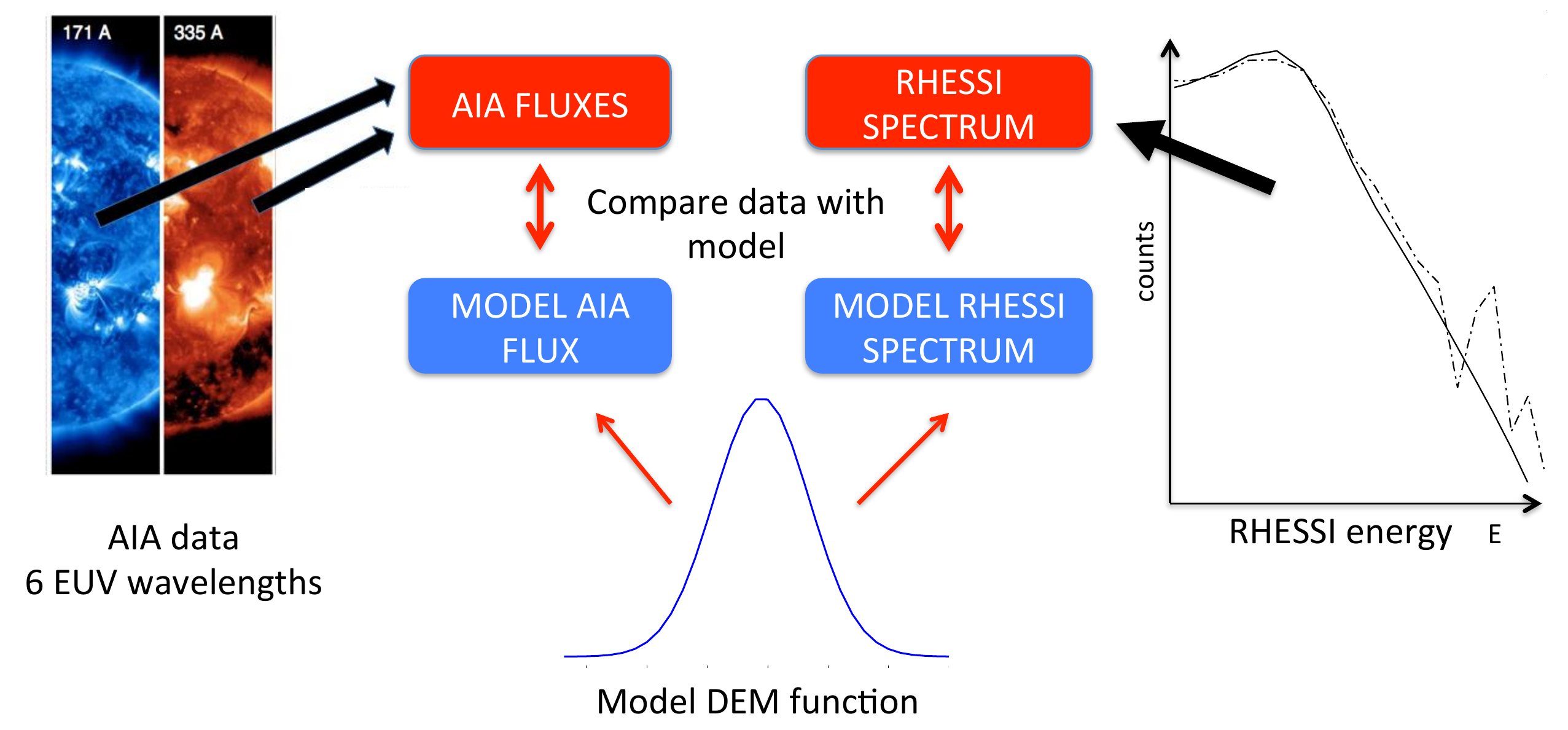}
\caption{Schematic of the joint differential emission measure fitting procedure. AIA flux data from the six optically thin EUV channels is compared to the reconstructed flux from a model DEM function folded through the AIA temperature response function. Simultaneously, the RHESSI count spectrum is compared to a model count spectrum obtained by calculating the expected photon spectrum that results from the model DEM function, and utilizing the RHESSI response function to convert from a photon spectrum to a count spectrum.}
\label{joint_fit_summary}
\end{center}
\end{figure*}

Several techniques are currently available to determine the differential emission measure (DEM) of hot solar plasma. The core challenge of determining a DEM is that it is an inverse problem which generally lacks a unique solution \cite[see for example][]{2014SoPh..tmp...31R}. Therefore constraining the solution space using multiple instruments and physical understanding is important. 

Most work with DEM distributions to date has been done for non-transient active region sources, using various methods, including regularization and Maximum Entropy, (\citealt{fludra1986}; \citealt{hannah2012}), fitting with cubic spline functions (\citealt{brosius1996}), Markov chain Monte Carlo (MCMC) (\citealt{kashyap1998}), and forward fitting of single and multi-Gaussian functions \citep[e.g.][]{aschwanden2001, aschwanden2005, aschwanden2011}. DEM calculations for flares are less common, but can also be found, using forward-fitting and MEM \citep{mctiernan1999}, regularization \citep{prato2006, 2013ApJ...771..104F}, maximum likelihood \citep{kepa2008}, and Monte Carlo methods (e.g., \citealt{reale2001})

In this paper we chose to apply a forward-fitting approach based upon that used by \citet{aschwanden2011} on AIA images in the six optically thin wavelengths; 94\AA, 131\AA, 171\AA, 193\AA, 211\AA\ and 335\AA. However in this work we expand the fitting procedure to consider additional functions and also to simultaneously fit RHESSI spectra (see Figure \ref{joint_fit_summary}). 

To estimate the DEM, fluxes for each wavelength are obtained from co-temporal AIA images. In order to study the same plasma with both AIA and RHESSI first RHESSI images are generated at the event peak X-ray emission in the 6-12 keV channel using the CLEAN algorithm \citep{2002SoPh..210...61H}. AIA data is averaged over the image integration time for the RHESSI image as well as over the area within the 50\% contour of the RHESSI flare image. These intervals are listed in Table \ref{fit_table}. Sun-integrated hard X-ray spectra for that same period were also generated, with appropriate background subtraction.

As each AIA wavelength is sensitive to plasma at different temperatures (see Figure \ref{response_curves}), each provides an additional data point for use in the DEM forward fit. For a given model DEM, the flux expected in each AIA wavelength may be calculated by multiplying this DEM distribution by the instrument response function for the given wavelength. Here we use version 2 of the AIA temperature response function obtained from the SolarSoftWare (SSW) routine \verb|aia_get_response.pro|, including the \verb|chiantifix| and \verb|evenorm| correction factors. The expected flux is then compared with the real observed flux from AIA. The best fit is found by searching the parameter space (e.g. for a Gaussian DEM model peak temperature $T_p$ and Gaussian width $\sigma$) and finding the combination of parameters which produces most accurately the flux values in the observed images. Simultaneously, the goodness of fit for each iteration of model parameters to the RHESSI count spectrum is determined as described in Section \ref{fit_rhessi}. 

In this paper, we explore two DEM models. Firstly, we investigate the Gaussian model suggested by \citet{aschwanden2011}, which is characterised by a peak temperature $T_p$ and a Gaussian width factor $\sigma$, and can be written:

\begin{equation}
DEM = DEM_{0} \exp{\left(-\frac{(\log{T} - \log{T_p})^2}{2 \sigma^2}\right)},
\label{dem_equation}
\end{equation}

where $DEM_{0}$ is the peak emission measure of the distribution.

Secondly, to model a uniform emission measure we choose an Epstein function \citep[see][for previous examples of use of this function]{1995SoPh..159..399N, 2007A&A...461..1149P, 2009A&A...503..569I}, which may be written,

\begin{equation}
DEM = DEM_{0} sech^2 \left( \left[ \frac{\log{T} - \log{T_p}}{\sigma} \right]^n \right)
\label{epstein_eqn}
\end{equation}

\begin{figure}[h]
\begin{center}
\includegraphics[width=8cm]{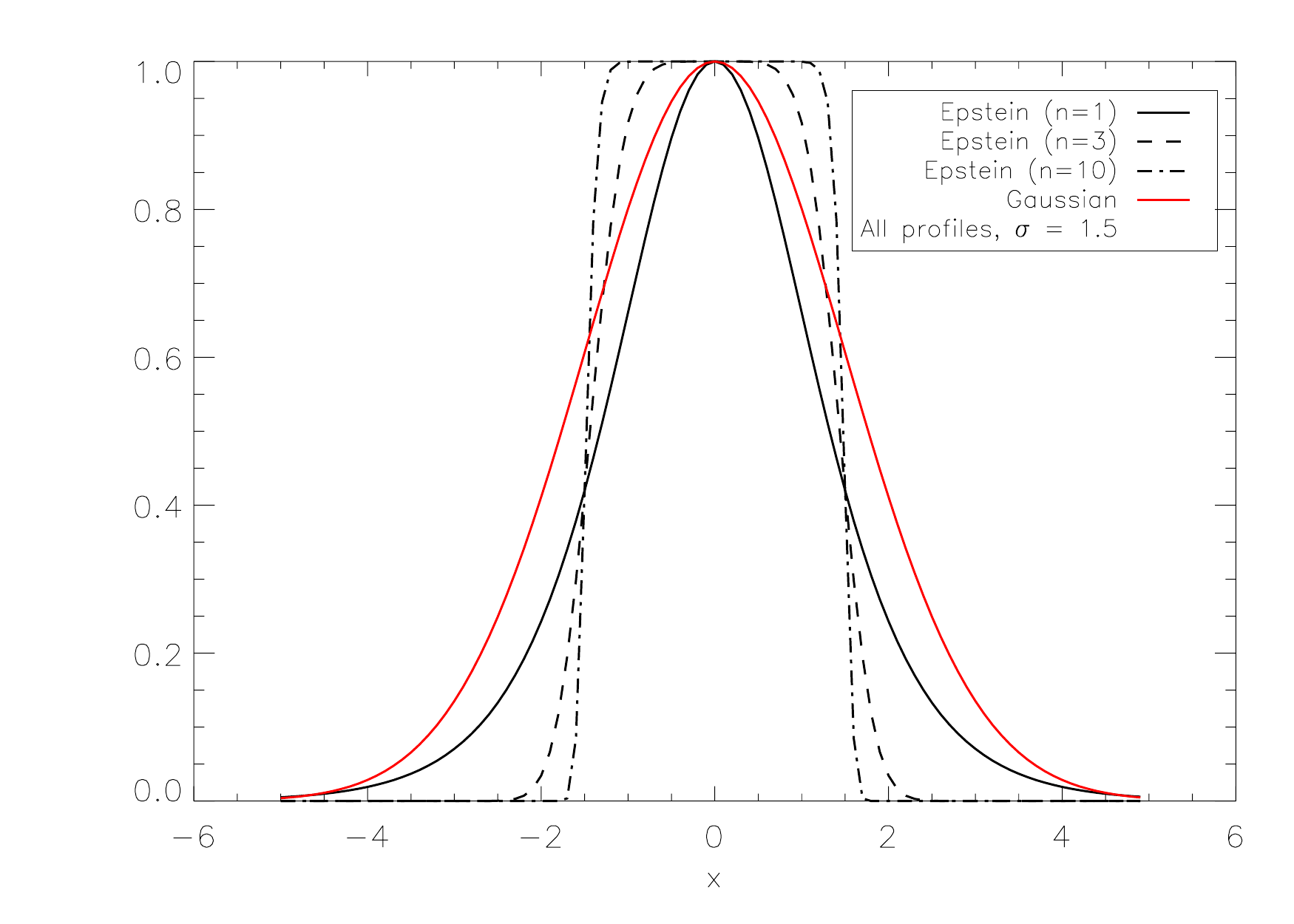}
\caption{A normalised Epstein function centred at $x$ = 0, with steepness parameters of $n$ = 1 (solid line), 3 (dashed line) and 10 (dot-dashed line). For comparison a Gaussian centred at $x$ = 0 is given by the red line. In all cases, width factor $\sigma$ = 1.5. }
\label{epstein_illustration}
\end{center}
\end{figure}

where $\sigma$ and $DEM_{0}$ are width and amplitude parameters as before, $T_p$ remains a temperature parameter denoting the center of the distribution, and $n$ is a steepness parameter. For both models, the parameter search space consists of 40 $T$ values equally distributed in log space between $\log T$ = 5.5 and $\log T$ = 7.5, and 75 $\sigma$ values equally spaced in the range 0.05 - 0.8. At each $T$ and $\sigma$, the amplitude parameter $DEM_0$ is estimated by minimising the difference between the observed and the modelled AIA flux.

This distribution is convenient as it allows a smooth variation between a classical Epstein profile, where $n = 1$, to a boxcar function where $n \to \infty$ (see Figure \ref{epstein_illustration}). Throughout this paper we assume a boxcar approximation by setting $n = 10$. This choice tests a simple case for the functional form of the DEM, which is that an entirely uniform emission profile may adequately fit the combined AIA and RHESSI data, between low and high cut-off temperatures. 
The contention is not that such a simple DEM distribution represents the actual emission structure of these events, but that it serves as a useful first-order approximation. This hypothesis gives us insight into whether the observations require more complex forms of the DEM \citep[e.g.][]{2012ApJ...759..141W, 2013ApJ...770..116W} and by extension additional parameters. Also, this function is able to fall off steeply at high $T$, an important consideration given previous studies of RHESSI spectra which have adequately fit high temperature plasma with power-law or exponentially decaying emission \citep[e.g.][]{hannah2008}. For convenience we further define $T_{max} = T_p + \sigma$ and $T_{min} = T_p - \sigma$, as these parameters provide a more intuitive description of a uniform emission profile than $T_p$ and $\sigma$.

Even for relatively small events such as microflares, saturation of AIA images can occur. Saturation is characterized by bleeding of flux from one pixel into neighbouring pixels, the result being that the true flux information is spread over a substantial image portion. Significant image saturation can lead to inaccurate DEM fit results. To account for this, each microflare was examined during the chosen fitting interval to find the fraction of pixels under the 50\% RHESSI contour in each AIA image where the data number exceeded 1.6$\times10^4$. This level corresponds approximately to the saturation level of AIA \citep{2011ApJ...743L..27R}. 

Only two microflares showed any significant saturation, defined here as $>$ 10\% of the relevant pixels. Event $\#$4 (2011 July 16) contained one saturated frame and event $\#10$ (2012 September 27) contained two saturated frames. These frames were removed from the analysis, meaning that the AIA flux data is averaged over fewer frames for these two events.

Another issue is that the temperature response functions of SDO/AIA are associated with significant uncertainty that affects the best-fit DEM and subsequent energy estimates. These uncertainties remain the subject of active study \citep[e.g.][]{aschwanden2011}. In particular, the 94\AA\ and 131\AA\ channels are known to be affected at low temperatures by lines not currently accounted for in the \verb|chianti| database. To account for this, we adopt a systematic uncertainty of 25\% for the flux detected in each AIA channel as suggested by \citet{2012SoPh..275...41B,2012ApJS..203...25G}.

\subsection{Fitting RHESSI spectra}
\label{fit_rhessi}

RHESSI spectra are critical in order to provide high temperature constraints to the DEM model. Fits to the RHESSI count spectra were made using the OSPEX analysis package available through SSW. New fitting routines were developed for this investigation, \verb|multi_therm_gauss| and \verb|multi_therm_epstein|, and have been made freely available within OSPEX, where their functionality is described in detail\footnote{OSPEX documentation is available online at \url{http://hesperia.gsfc.nasa.gov/rhessi2/}}. The fundamental feature of these routines is that, for a given set of DEM function parameters, the expected photon spectrum is generated by assuming bremsstrahlung continuum emission. Subsequently, the RHESSI detector response matrix (or DRM) is used to predict the count flux that would be observed by RHESSI. This flux is compared with the observed flux in each energy bin within the fitting range, and a $\chi^2$ minimization is performed.

Typically, for microflares RHESSI detects significant emission only in the range $E \leq 15$ keV. In this study we fit background-subtracted RHESSI spectra, where the fitting time interval spans 60 seconds. This length of time integration is necessary for small events such as microflares in order to accumulate sufficient counts to generate an accurate spectrum. Table \ref{fit_table} lists the fit and background intervals of each event. The fitting energy range is restricted to 4-8 keV since above 8 keV non-thermal effects can be important. Additionally, for consistency the DEM fits are performed on data from detector 1 of RHESSI in all cases. For additional simplicity, only the model spectrum derived directly from the DEM model is fit to the RHESSI count spectrum; other common corrections, such as albedo effects, and modification of the elemental abundances are neglected in order to reduce the number of free parameters, although it should be noted that including such small corrections would likely improve the RHESSI goodness-of-fit. To account for this we allow for greater systematic uncertainty associated with the RHESSI data points, adopting a value of 4\% throughout this work. This includes a standard 2\% level uncertainty corresponding to uncertainty in the base RHESSI response function, combined with an additional 2\% estimated uncertainty accounting for the small corrections listed above.

Throughout this paper we refer to the reduced $\chi^2$ values for AIA or RHESSI data alone defined by,

\begin{equation}
\chi_{AIA}^2 =  \frac{1}{N_{AIA} - n -1} \sum_{AIA} \frac{(f_{o} - f_{m})^2}{\sigma_{AIA}^2}
\end{equation}

and,

\begin{equation}
\chi_{HSI}^2 = \frac{1}{N_{HSI} - n -1}  \sum_{HSI} \frac{(f_{o} - f_{m})^2}{\sigma_{HSI}^2} 
\end{equation}

where $f_{o}$ is observed flux, $f_{m}$ is the expected model flux, $\sigma$ denotes the variance, $n$ is the number of free parameters and $N_{AIA}$ and $N_{HSI}$ represent the number of data points available for AIA and RHESSI respectively. For both models used in this paper, the free parameters are $\log T_p$, $\sigma$ and $EM_0$ (see Equations \ref{dem_equation},\ref{epstein_eqn}), and hence the number of free parameters $n$, is 3.

The combined reduced $\chi^2$ is denoted by $\chi_c^2$ and is obtained as follows,

\begin{equation}
\chi_{total}^2 = \sum_{AIA} \frac{(f_{o} - f_{m})^2}{\sigma_{AIA}^2} + \sum_{HSI}  \frac{(f_{o} - f_{m})^2}{\sigma_{HSI}^2} \\
\label{chisq_eqn0}
\end{equation}

and,

\begin{equation}
\chi_c^2 = \frac{\chi_{total}^2}{N-n-1}
\label{chisq_eqn}
\end{equation}

where $N = N_{AIA} + N_{HSI}$ is the total number of data points. In this study, $N_{HSI}$ is always larger than $N_{AIA}$, which equals 6 in all cases, whereas over the 4-8 keV energy range $N_{HSI}$ = 12. The consequence is that the RHESSI data contribute more strongly to $\chi^2_c$.

\section{Results}
\label{results}

\begin{figure*}
\begin{center}
\includegraphics[width=15cm]{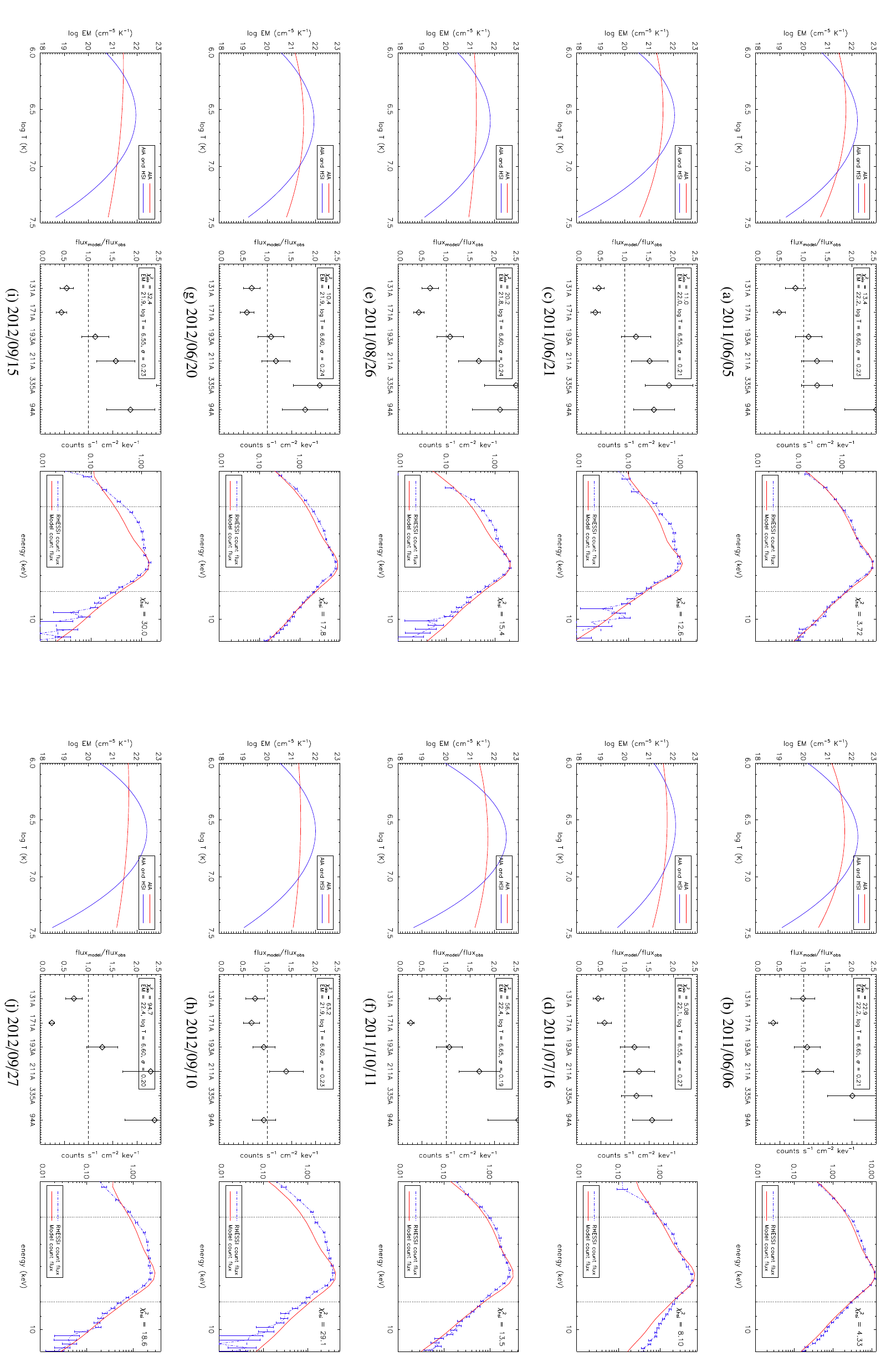}
\caption{Results of joint forward-fitting of a Gaussian DEM profile to AIA fluxes and RHESSI count spectra. The events are ordered sequentially according to the event date. For each sub-image, the leftmost panels show the combined best-fit DEM as a function of $\log T$ (blue line). For comparison the DEM which best fits the AIA data alone (red line) is also shown. The centre panels show the ratio of the best-fit model AIA flux to the observed flux, while the rightmost panels show the comparison between the model count spectrum and the observed RHESSI spectrum.}
\label{gauss_subfig}
\end{center}
\end{figure*}

\begin{figure*}
\begin{center}
\includegraphics[width=15cm]{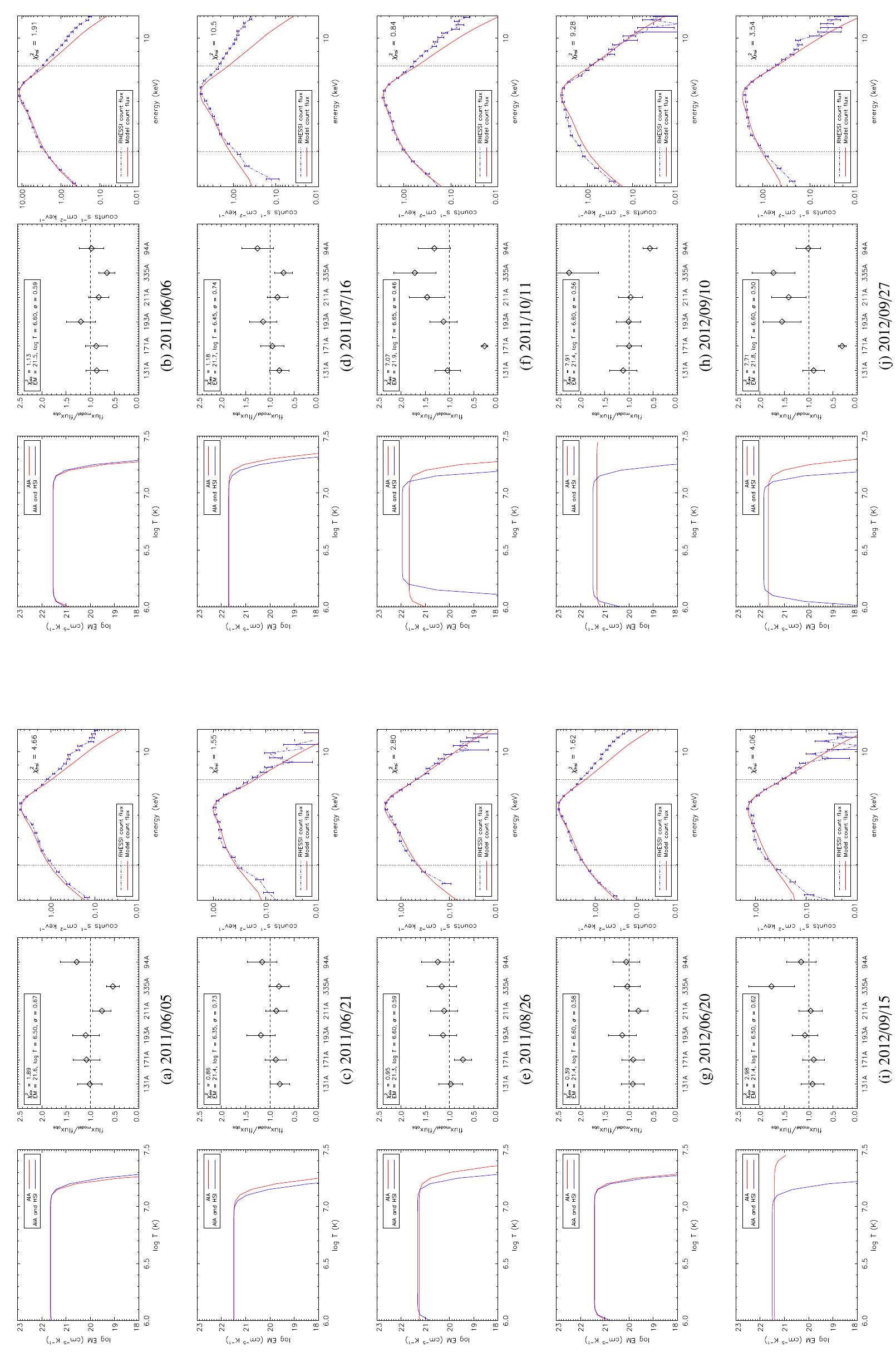}
\caption{Results of joint forward-fitting of an Epstein DEM profile to AIA fluxes and RHESSI count spectra. The events are ordered sequentially according to the event date. For each sub-image, the leftmost panels show the combined best-fit DEM as a function of $\log T$ (blue). For comparison the DEMs which best fit the AIA data alone (red line) is also shown. The centre panels show the ratio of the best-fit model AIA flux to the observed flux, while the rightmost panels show the comparison between the model count spectrum and the observed RHESSI spectrum.}
\label{epstein_subfig}
\end{center}
\end{figure*}

We analyse the full set of events listed in Table \ref{fit_table} and Table \ref{energy_table}, in each case carrying out joint fitting of the AIA flux data and the RHESSI count spectrum as illustrated in Figure \ref{joint_fit_summary} and described in Section \ref{fitting_dems}. Both the Gaussian and Epstein profiles are used, and the results of joint fitting using each model are displayed in Figure \ref{gauss_subfig} and Figure \ref{epstein_subfig}, respectively.

\begin{savenotes}
\begin{table*}

\caption{Summary of best-fit results for selected microflares.}
\centering
\scalebox{1.1}{
\begin{tabular}{ccccc|cc|ccc}
\tableline

\# & Event Date & GOES & GOES & RHESSI & \multicolumn{2}{c}{Gaussian} & \multicolumn{3}{c}{Epstein} \\
& & Class & Peak Time & fit time (UT) & $\chi^2_c$ & [$\log$ T,$\sigma$] & $\chi^2_c$ &  [$\log$ T,$\sigma$] & [$T_{min}$, $T_{max}$] \\
& & & (UT) & & & & & & (MK)  \\

\tableline

1 & 2011/06/05 & B3.5 & 02:14 & 02:13 - 02:14 & 4.9 &   6.60, 0.23   & 3.2 & 6.5, 0.67 & 0.7, 14.8 \\
2 & 2011/06/06 & B6.7 & 13:20 & 13:19 - 13:20 & 7.2 & 6.65, 0.21     & 1.4 & 6.6, 0.59 & 1.0, 15.4  \\
3 & 2011/06/21 & B2.8 & 18:22 & 18:22 - 18:23 & 9.7 & 6.55, 0.21 & 1.1 & 6.35, 0.73 & 0.4, 9.5 \\
4 & 2011/07/16 & B6.2 & 17:04 & 17:02 - 17:03 & 5.9 & 6.55, 0.27   & 6.5 & 6.45, 0.74  & 0.5, 15.4 \\ 
5 & 2011/08/26 & B4.2 & 20:53 & 20:53 - 20:54 & 13.3 & 6.60, 0.24   &  1.9 & 6.60, 0.59 & 1.0, 15.4  \\
6 & 2011/10/11 & B5.8 & 00:35 & 00:35 - 00:36 & 19.4 & 6.65, 0.19   & 2.0 & 6.65, 0.46 & 1.5, 12.9  \\
7 & 2012/06/20 & B7.8 & 15:53 & 15:48 - 15:49 & 12.8 & 6.60, 0.24   & 1.1 & 6.6, 0.58 & 1.0, 15.0 \\
8 & 2012/09/10 & B8.7 & 07:24 & 07:22 - 07:23 & 30.1 & 6.60, 0.23   &  7.2 & 6.60, 0.56 & 1.1, 14.5  \\
9 & 2012/09/15 & B2.9 & 22:44 & 22:44 - 22:45 & 24.5 & 6.55, 0.23   & 3.0 & 6.50, 0.62 & 0.8, 13.2  \\
10 & 2012/09/27 & B7.5 & 06:57 & 06:56 - 06:57 & 30.1 & 6.60, 0.20   & 3.7 & 6.60, 0.50 & 1.3, 12.6 \\

\tableline
\label{fit_table}
\end{tabular}
}
\end{table*}
\end{savenotes}

\begin{savenotes}
\begin{table*}

\caption{Summary of energy estimates for selected microflares.}
\centering
\scalebox{1.1}{
\begin{tabular}{cccccccccccccc}
\tableline

\# &  $dL/dt$ & $t_{dur}$ & $L_{rad}$ & \multicolumn{4}{|c|}{multi + thick$^{(1)}$}  &  \multicolumn{4}{|c}{iso + thick$^{(2)}$}  \\

& ($10^{24}$ erg s$^{-1}$) & (s) & ($10^{27}$ ergs) & $U_{th}$ & $U_{nth}$ & $E_c$ & $U_{nth} / U_{th}$ & $U_{th}$ & $U_{nth}$ & $E_c$ & $U_{nth} / U_{th}$  \\ 
& & & & (10$^{29}$ ergs) & (10$^{29}$ ergs) & (keV) &  & (10$^{29}$ ergs) & (10$^{29}$ ergs) & (keV) & \\

\tableline

1 &  10.1   & 180   & 1.8  & 0.49 & 0.18 & 14.3 & 0.37 & 0.45 & 0.39 & 11.8 & 0.86 \\
2 & 20.4  &  420  &  8.6 & 1.5 & 0.81 & 12.3 & 0.54 & 1.5 & 1.5 & 11.1 & 1.00 \\
3 & 7.1 & 300 & 2.1 & 0.34 & 0.09 & 11.2 & 0.26 & 0.36 & 0.11 & 11.4 & 0.30 \\
4 & - & - & - & - & - & - & - & 0.45 & 0.19 & 11.5 & 0.42 \\
5 & 5.4 & 480 & 2.6 & 0.39 & - & - & $<$0.01 & 0.42 & 0.43 & 11.8& 1.02 \\
6 & 21.6 & 300 & 6.5 & 0.72 & 0.14 & 15.2 & 0.19 & 0.71 & 0.66 & 9.4 & 0.93 \\
7 & 16.4 & 900 & 14.7 & 1.3 & 0.47 & 14.3 & 0.36 & 1.34 & 1.23 & 11.8 & 0.92 \\
8 & - & - & - & - & - & - & - & 0.95 & 0.35 & 9.6 & 0.37 \\
9 & 8.2 & 180 & 1.5 & 0.47 & 0.14 & 10.5 & 0.30 & 0.55 & 0.07 & 12.7 & 0.13 \\
10 & 17.0 & 240 & 4.0 & 0.64 & 0.07 & 9.9 & 0.11 & 0.63 & 0.32 & 5.2 & 0.51 \\

\tableline
\label{energy_table}
\end{tabular}
}
\caption*{\textbf{Notes}: \\ $dL/dt$ and $L_{limit}$ are derived using Epstein DEM profile in all cases. \\ (1) Thermal and non-thermal energy calculated by using the best-fit multi-thermal model and fitting the remaining high-energy emission with a thick target brehmsstrahlung model. \\ (2) Thermal and non-thermal energy calculated using a traditional isothermal-plus-thick target spectral model.}
\end{table*}
\end{savenotes}

Figure \ref{gauss_subfig} shows the performance of the Gaussian model, where three panels are plotted for each microflare. The leftmost panel in each case shows two DEM models. The first is a reference best-fit where only the AIA flux data has been used (red line), and the second is a combined fit utilising both AIA and RHESSI data (blue line). The center panel shows the ratio between the combined best-fit modelled AIA flux and the actual observed flux in each channel. Finally, the right panel shows the comparison between the combined best-fit model count flux, and the actual count flux observed by RHESSI.

The $\chi^2_c$ values for each microflare (see Table \ref{fit_table}) indicate that these events are not well-represented using a single Gaussian DEM model, with $\chi^2_c \approx 5$ for event \#1 the best value obtained. Figure \ref{gauss_subfig} reveals that the combined best-fit DEM is consistently shifted to favour a narrower distribution when compared to best-fits obtained using AIA data only. However, this combined fit tends to provide a poor reproduction of both the AIA fluxes and the RHESSI spectra. Hence this model appears to be inappropriate for constraining both high and low temperature plasma simultaneously. This is discussed further in Section \ref{case_study}.

The result of applying the Epstein model to the events is presented in Figure \ref{epstein_subfig}. Here we observe mixed results. For events $\#2$, $\#3$, $\#5$ and $\#7$, the model can be said to fully describe the data, with $\chi^2_c < 2$. These values may be re-cast in terms of the probability value (or $p$-value), where values of $p$ below a set threshold (e.g. 0.05 for a 95\% significance level or 0.01 for a 99\% level) indicate that the model should be rejected as a full description of the data. The $\chi^2_c$ values for these events correspond to probability values of $p_{\#2} = $0.14, $p_{\#3} = $0.35, $p_{\#5} = $0.02 and $p_{\#7} = $0.35 respectively.

Of the fully-fit events, a visual inspection of the RHESSI spectra in Figure \ref{epstein_subfig} reveals that microflares $\#3$ and $\#5$ are in fact fully fit by the multi-thermal model at energies above 8 keV. This implies that the non-thermal component of emission is either absent in these events or is sufficiently weak to be entirely masked by the thermal emission. For events $\#2$ and $\#7$, there is, by contrast, evidence of a break above 8 keV to a power-law distribution, implying the presence of significant non-thermal energy.

For the remaining events, four have a $\chi^2_c$ value between 2.0 and 4.0 (events $\#1$, $\#6$, $\#9$ and $\#10$). For these, the chosen model does not fully describe the data. However, Figure \ref{epstein_subfig} shows that the form of the RHESSI spectrum is reasonably well reproduced for these events, as is the AIA flux at the 131\AA, 193\AA, 211\AA\ and 94\AA\ wavelengths. The 171\AA\ and 335\AA\ emission for these events is the least well reconstructed. For example, in event $\#9$ the predicted 335\AA\ flux is $\approx$ 1.75 that of the observed flux, while in event $\#10$ the 171\AA\ flux is underestimated by a factor of 3. Since both of these channels have a substantial temperature response component below $\log T = 6.0$, one explanation is that the model DEM function is not fully accounting for this lower temperature emission.

The remaining two events, $\#4$ and $\#8$, measure $\chi^2_c=6.5$ and $\chi^2_c=7.2$ respectively. Hence the simple uniform DEM distribution applied here is insufficient to describe these events. Since a single Gaussian DEM distribution also proved inadequate, the use of more complex DEM distributions in these cases is justified.

\subsection{A case study: 2011 June 21}
\label{case_study}

\begin{figure}
\begin{center}
\includegraphics[width=8cm]{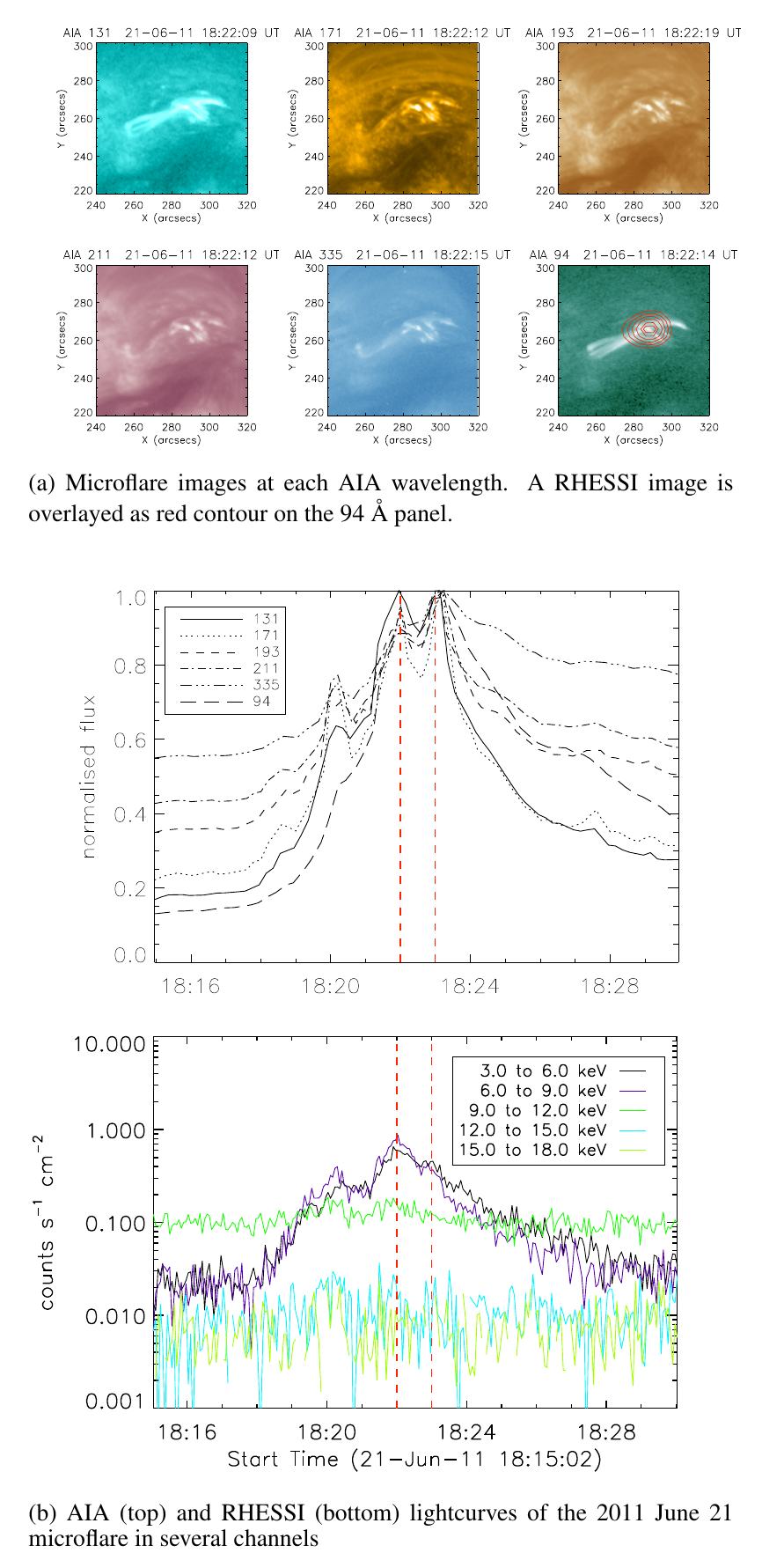}
\caption{Summary information for the 2011 June 21 microflare from 18:20 UT. \textit{a)} AIA images of the microflare at each EUV wavelength, with a reconstructed RHESSI image overlayed on the 94 \AA\ image as a red contour. The RHESSI image was constructed in the 6-12 keV range using the CLEAN algorithm and shows emission above the 50\% level. \textit{b)} AIA lightcurves of the microflare region and immediate surrounds, and RHESSI lightcurve of the event displayed in a range of energy bands.}
\label{event3}
\end{center}
\end{figure}

In order to explore these results in detail, an in-depth study of event \#3 is presented here. This microflare originated from active region AR11236 and occurred at approximately 18:20 UT, as shown in Figure \ref{event3}. Figure \ref{event3}b contains both AIA lightcurves in each waveband and RHESSI lightcurves in a number of X-ray channels. The fitting interval used is denoted by the vertical dashed lines. Figure \ref{event3}a illustrates the spatial extent of the microflare in both AIA and RHESSI, with the RHESSI image shown as a contour on the 94\AA\ panel from AIA.

\subsubsection{Fitting AIA fluxes in isolation}
\label{example_aia_only}

\begin{figure}
\begin{center}
\includegraphics[width=8cm]{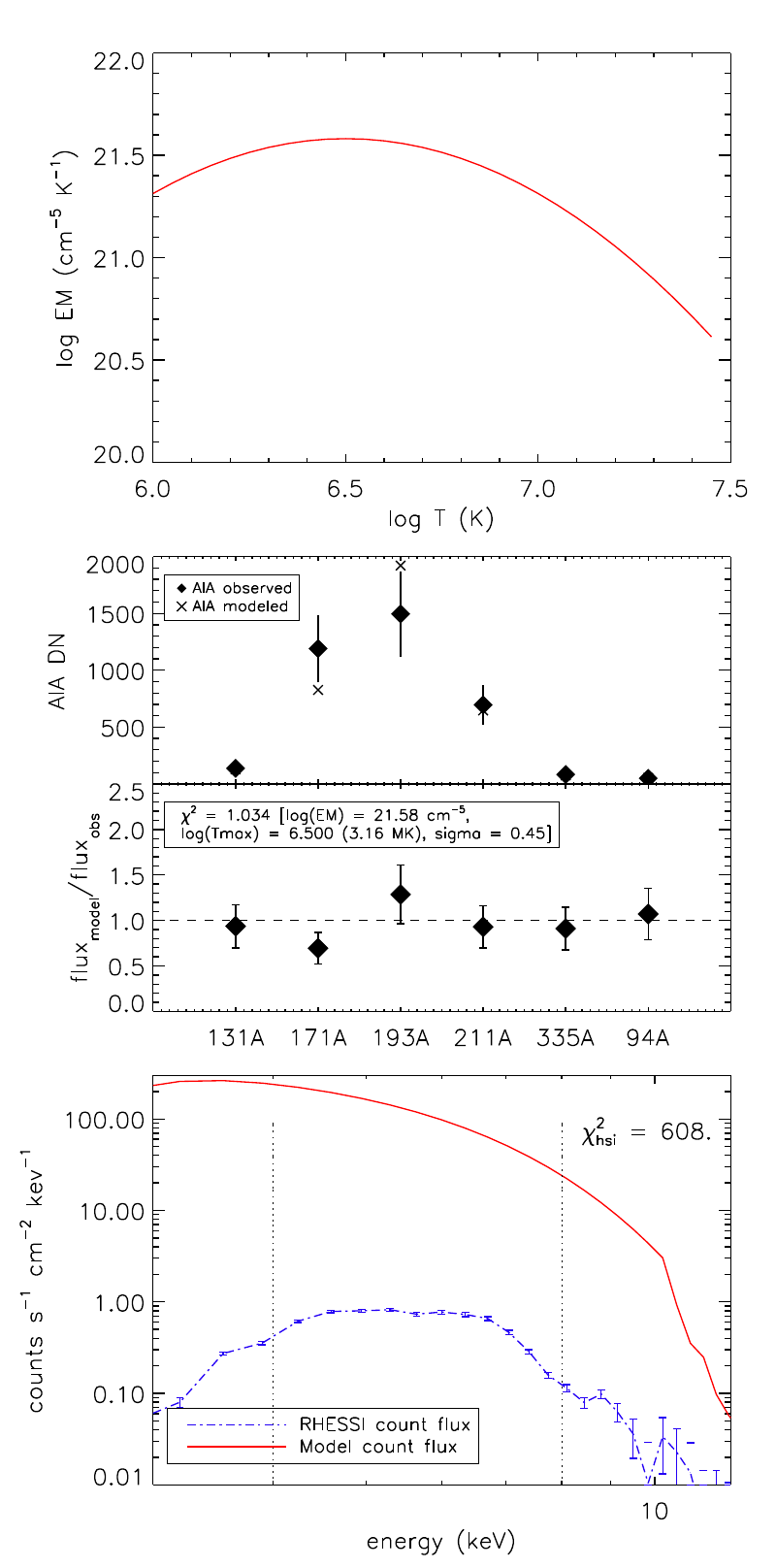}
\caption{Results of applying a best-fit Gaussian DEM function to the AIA flux data alone for the 2011 June 21 microflare. Top panel: The best fit model DEM. Centre panel: AIA flux comparison, where the actual observed flux is denoted by the diamonds and the model flux associated with the best-fit Gaussian DEM is denoted by the crosses. The ratio of the model flux to the observed flux is also shown for each AIA channel. Bottom panel: Comparison of the model RHESSI count spectrum that results from the best-fit DEM (solid line) to the actual energy spectrum observed by RHESSI. The observed spectrum was obtained between 18:22 and 18:23 UT on 2011 June 21 (dashed line).}
\label{aia_summ}
\end{center}
\end{figure}

Many previous studies of differential emission measure have relied on a single instrument only. To understand the contrast between this and joint analysis we constrain the DEM for event $\#3$ using data from AIA only. For the popular Gaussian DEM model, the results of this process are displayed in Figure \ref{aia_summ}. 

The top panel displays the best-fit DEM, while the center panel compares the observed AIA flux data with the model flux in each channel derived from the best-fit Gaussian model. These panels show that the flux from five of the six channels is reproduced to within the error bars, while the remaining channel - 171 \AA\ - has a slightly larger discrepancy. In this case $\chi^2_{AIA} \approx 1.0$, corresponding to a good fit. The Gaussian parameters are:  $\log EM_0 = $ 21.58 $cm^{-5}K^{-1}$, $\sigma = 0.45$, and $\log T_{p} = 6.50$, equivalent to $\approx$ 3.2 MK.

However, although the best-fit function is able to reproduce the observed fluxes in each AIA channel, in the bottom panel of Figure \ref{aia_summ} the resulting RHESSI count spectrum is compared with the observed background-subtracted energy spectrum. A difference of approximately two orders of magnitude is evident.

Hence, although an excellent fit to the  AIA data, this DEM dramatically overestimates the emission at high temperatures. This is a limitation of the Gaussian function, which is unable to decrease sufficiently quickly at high $T$ when $\sigma$ is large (e.g. $\sigma > 0.2$), and is already too broad to correctly fit the low temperature plasma.

\begin{figure}
\begin{center}
\includegraphics[width=8cm]{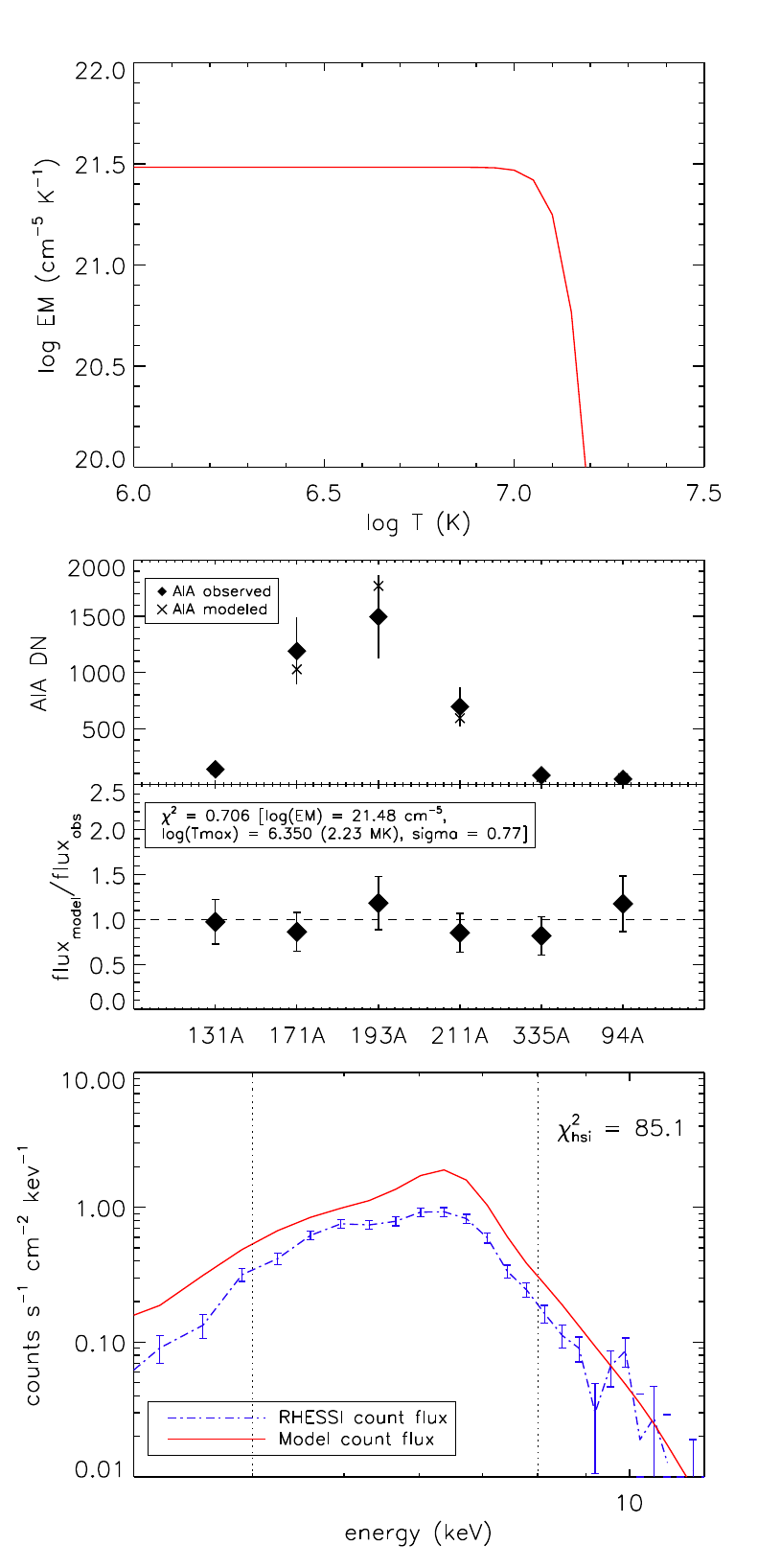}
\caption{Results of applying a best-fit Epstein DEM function to the AIA flux data alone for the 2011 June 21 microflare. Top panel: The best-fit model DEM. Centre panel: AIA flux comparison, where the actual observed flux is denoted by the diamonds and the model flux associated with the best-fit DEM is denoted by the crosses. The ratio of the model flux to the observed flux is also shown for each AIA channel. Here $\chi^2_{AIA} = 0.7$. Bottom panel: Comparison of the model RHESSI count spectrum that results from the best-fit DEM (solid line) to the actual energy spectrum observed by RHESSI (dashed line).}
\label{aia_summ_epstein}
\end{center}
\end{figure}

We similarly apply the Epstein profile to the AIA data for event $\#3$, without utilising RHESSI. The result of this fit is shown in Figure \ref{aia_summ_epstein}.

Here the DEM function again provides an excellent fit to the AIA flux data, with $\chi^2_{AIA} \approx 0.7$ and all six channel fluxes reproduced within the error bar. Moreover, the bottom panel in this case shows that the expected count flux in RHESSI (red line) is of the same order as the observed flux (blue, dashed line). However, the best-fit model remains unable to truly reproduce the RHESSI spectrum. 

The fact that both the Gaussian and Epstein models can be successfully fit to the AIA flux shows that AIA alone provides cannot distinguish between the models. Furthermore, the bottom panels of Figure \ref{aia_summ} and Figure \ref{aia_summ_epstein} imply that, regardless of the chosen model, AIA data is insufficient to properly constrain emission at high temperatures. 

This is further confirmed by carrying out AIA-only fits on each of the sampled events. For each microflare, the RHESSI counts estimated from the Gaussian model exceed the actual observed emission by at least one order of magnitude, usually two, similar to that observed in Figure \ref{aia_summ}. The Epstein profile by contrast provides RHESSI count estimates of the same order as the observed spectrum 80\% of the time, the exceptions being events $\#8$ and $\#9$. However, the $\chi^2_{HSI}$ values remain poor (generally in excess of 10), indicating that the AIA data is insufficient to determine the high temperature cutoff correctly.

\subsubsection{Combining AIA and RHESSI in a joint fit}
\label{example_combo}

For event $\#3$ we examine the joint AIA and RHESSI fits for both the Gaussian and Epstein models. The results of the Gaussian model are shown in Figure \ref{gauss_subfig}g, and the Epstein model in Figure \ref{epstein_subfig}e.

The result of the combined fitting of the Gaussian model is significantly different from the AIA-only fit. The model RHESSI spectrum is closer (albeit still a poor fit, with $\chi^2_{HSI} = 12.6$) to the observed count spectrum, as a result of the best-fit model being much narrower ($\sigma = 0.21$). This narrower Gaussian means that the contributions to the total emission reduce much more quickly in the high $T$ range, thus avoiding the unphysical excess contributions discussed in Section \ref{example_aia_only}. However, the consequence of this is that the narrow Gaussian is no longer able to accurately reproduce the observed AIA fluxes, as evidenced in the center panel of Figure \ref{gauss_subfig}g. Hence, the model fits neither the AIA or RHESSI data well, and the overall goodness of fit, $\chi^2_c = 9.7$, is unacceptable.

The Epstein profile in contrast is able to fit the AIA and RHESSI data simultaneously, as Figure \ref{epstein_subfig}e shows. Five of the six AIA fluxes are reproduced within the error bars, and the RHESSI spectrum is well reproduced, with the result that $\chi^2_c = 1.1$.

For event $\#3$, a constant DEM distribution as a function of $T$, combined with a steep fall-off at a high temperature, provides a good fit to the joint dataset, particularly considering that higher order corrections to the RHESSI spectrum were neglected (see Section~\ref{fit_rhessi}). By comparison, joint fitting using a Gaussian model was unable to simultaneously fit both the AIA fluxes and the RHESSI spectrum with a single consistent function. Both Figure~\ref{gauss_subfig} and Table~\ref{fit_table} show that this limitation of the Gaussian function was replicated for all of the microflares in the sample, with $\chi^2_c > 4.0$ in all cases, and generally much higher. Hence this model cannot be used to understand plasma emission over a wide temperature range.

By contrast, the results achieved with the Epstein model consistently improve upon those obtained with the Gaussian model, with $\chi^2_c < 2.0$ in four events. In four additional events, $2.0 < \chi^2_c < 4.0$. Hence, based on the RHESSI and AIA data, most of the studied microflares may be adequately fit by a uniform differential emission measure profile with a steep high temperature cutoff, implying that any finer structure in the DEM cannot be resolved by these instruments.

\section{Microflare energetics}
\label{energy}
\subsection{Radiative losses}

Given a well-constrained DEM for a plasma, it is possible to estimate the radiated energy loss rate as follows \citep[e.g.][]{2005psci.book.....A, 2012ApJS..202...11R}.

\begin{equation}
\frac{dL}{dt} = \int DEM(T) \times \Lambda (T) \ dT
\end{equation}

In this Equation $DEM(T)$ is the amount of emission - multiplied by event area $A$ and hence in units of cm$^{-3} K^{-1}$, and $\Lambda (T)$ represents the radiative loss function, which has been investigated by many authors \citep[e.g.][]{1978ApJ...220..643R, 1981A&AS...45...11M, 1989ApJ...338.1176C, 2000ApJ...537..471M, 2008ApJ...682.1351K, 2011A&A...529A.103D}.

An estimate of $\Lambda (T)$ may be obtained from the \verb|chianti| database after choosing appropriate coronal abundances. Although this function is only weakly dependent on density \citep[e.g.][]{2005SoPh..227..231W}, the radiative losses vary by an order of magnitude as a function of temperature in the range $0.5 < T < 5$ MK.

The total radiated energy can be estimated by the integral of the radiative loss rate $dL/dt$ over the duration $t_{dur}$ of the event. However, for microflares it is difficult to resolve the behaviour of the DEM distribution as a function of time, given the need to integrate the RHESSI counts for up to 60 s in order to obtain a viable spectrum. Instead, we estimate the total radiated energy as $L_{rad}$ $\approx$ $dL/dt \times t_{dur}$, where $t_{dur}$ is the duration of the microflare estimated from the event lightcurves.

Estimates of $dL/dt$ and $L_{rad}$ are obtained for 8 of the 10 events, with events $\#4$ and $\#8$ discarded due to poor fits. We find that the estimated radiative loss rates range between $5 \times 10^{24}$ - $2 \times 10^{25}$ erg s$^{-1}$, while the total radiative losses $L_{rad}$ range between $2 \times 10^{27}$ -  $1.5 \times 10^{28}$ ergs, with a mean value of $\approx 5 \times 10^{27}$ ergs. The estimates for each event may be found in Table \ref{energy_table}.

This can be compared with estimates obtained from background-subtracted GOES data, where an isothermal plasma and coronal abundances are assumed \citep{2005SoPh..227..231W}. To estimate the background for each event, a featureless time interval prior to the event was selected and taken to be the background level. To estimate $L_{rad}$, the estimated radiative losses $dL/dt$ were integrated over the same duration as listed in Table \ref{energy_table}. We find that the GOES energy estimates are of the same order of magnitude as those obtained from the fitted DEM distribution. In general, the GOES-estimated energies are higher than the estimates obtained from the multi-thermal DEMs, by a factor $\approx$ 2.

\subsection{Thermal energy}

For an isothermal plasma at temperature $T$, the thermal energy is given by \citep[e.g.][]{hannah2008,2012ApJ...759...71E},

\begin{equation}
U_{th} = 3 n_e k_B T f V,
\end{equation} 

where $f$ is the filling factor, $n_e$ is the electron number density, $V$ is the plasma volume and $k_B$ is Boltzmann's constant. Since $n_e = \sqrt{EM/fV}$, this may be rewritten as

\begin{equation}
U_{th} = 3 k_B T \sqrt{EM f V}.
\label{u_th}
\end{equation}

A recent statistical study by \citet{hannah2008} analysed more than 25,000 microflares in this way, finding that their median thermal energy at the peak of the event was of the order $10^{28}$ ergs, with a $2\sigma$ range of $10^{26}$ - $10^{30}$ ergs. In this study, the median volume $V$ was estimated as $10^{27}$ cm$^{3}$ with a $2\sigma$ range of $5\times10^{25}$ - $2\times10^{28}$ cm$^{3}$, and the median emission measure was found to be $3\times10^{46}$ cm$^{-3}$ with range $4\times10^{45}$ - $2\times10^{47}$ cm$^{-3}$.

The thermal energy may also be estimated for a multi-thermal plasma. In this case, it is necessary to account for all the energy at all $T$. From Equation \ref{u_th} we may write,

\begin{equation}
%U_{th} = 3 k_B \int_{T_1}^{T_2} T \times \sqrt{EM(T) V} \ dT 
U_{th} = 3 k_B V^{1/2} \frac {\int_{T} DEM \times T dT}{EM^{1/2}} 
\label{multi_therm_energy_eqn}
\end{equation}

where $DEM = dEM/dT$, the differential emission measure (in cm$^{-3} K^{-1}$), $EM$ is the total emission measure and the assumption $f \approx 1$ has been made. The volume of the emitting material is estimated from the measured flare area as $V \approx A^{3/2}$. We apply Equation \ref{multi_therm_energy_eqn} to the 8 well-constrained microflares, utilising the best-fit Epstein DEM model in all cases. The total thermal energy $U_{th}$ is found to range between $3.9\times10^{28}$ - $1.5\times10^{29}$ (see Table \ref{energy_table} for individual estimates). These estimates are similar to the mean value established by \citet{hannah2008} with an isothermal approach. The volume estimates obtained for these events are also comparable with \citet{hannah2008}, ranging between $2.4\times10^{27}$ - $9.4\times10^{27}$ cm$^3$.

Comparison with the earlier derived values of $L_{rad}$ reveals an inconsistency. If a plasma is radiating simply, then the total radiated energy should not be less than the peak instantaneous thermal energy of the plasma. Indeed, during a recent study by \citet{2012ApJ...759...71E} of 38 large solar eruptive events, it was noted that in general the total radiated energy from an event exceeded the peak thermal energy by a factor of $\approx$ 3.  

For our measurements we find that $L_{rad} << U_{th}$, with the thermal energy exceeding the total radiated energy by around an order of magnitude. This occurs despite our estimates of the emission measure and volume being broadly consistent with those found by \citet{hannah2008}. 

The two main explanations for this energy inequality are that either thermal conduction losses are strong compared to radiative losses in the observed microflares, or that the filling factor $f$, assumed to be unity for these estimates, is in fact small. A recent study of large flares by \citet{2012ApJ...755...32G} has suggested that $ 0.05 < f < 0.78 $, however this may not hold for smaller events such as microflares \citep{1999ApJ...526..505M}. If we assume for our events the simplest relation between the thermal and radiated energy, assuming conductive losses to be small, that $L_{rad} \approx U_{th}$, then we can estimate lower limits on the filling factor $f$. For each event, we estimate $f$ based on this assumption, finding a mean value of $f \approx 5\times10^{-3}$. The lower limits on $f$ are similar to the estimates of the filling factor made by \citet{2011ApJ...736...75B} in another recent microflare study, indicative of the idea that microflares and large flares may have different characteristic filling factors. In reality however, the conductive losses in microflares are likely to be significant, particularly during the rise phase of the emission, accounting for at least some of the difference between $L_{rad}$ and $U_{th}$. To investigate this, we compared the contribution of non-thermal energy $U_{nth} / U_{th}$ (see Figure \ref{energy_balance} and Table \ref{energy_table}) in each event with $U_{th} / L_{rad}$. Weak non-thermal emission would suggest that more thermal conduction is taking place, as a mechanism is needed to evaporate hot material into the corona. Since this is not accounted for by $L_{rad}$, one might expect a larger value of $U_{th} / L_{rad}$ in these cases. However, perhaps due to the small sample size, no relationship was found between these two measurements, though it must be recognised that the estimates of $U_{nth}$ are only lower limits for each event.

\subsection{Non-thermal energy}
\label{nthermenergy}

A well-constrained model for the microflare thermal energy allows for a more accurate measurement of the non-thermal energy content by considering the remaining, high-energy, portion of the X-ray spectrum as non-thermal emission. Given a thick-target brehmsstrahlung model, the power in non-thermal electrons above the low-energy cutoff is given by \citep[e.g.][]{2013ApJ...771..104F},

\begin{equation}
%P_{nth}(>E_c) = 9.5\times10^{24} \gamma^{2} (\gamma - 1) \beta \left( \gamma - \frac{1}{2},\frac{3}{2} \right) I_0 E_c^{(1-\gamma)}
P_{nth}(>E_c) = \frac{\delta - 1}{\delta -2} \frac{F}{E_c} 
\label{ntherm_eqn}
\end{equation}

where $P_{nth}$ is in erg s$^{-1}$, $\delta$ is the power law spectral index of electrons, and $F$ is the number of electrons per second above the low energy cut-off $E_c$. 

As discussed by many authors \citep[see][]{2011SSRv..159..107H}, the low-energy cutoff $E_c$ can only be estimated, as it is masked by the thermal emission that dominates at lower energies. Hence, the values of $E_c$ used in Equation \ref{ntherm_eqn} should be considered upper limits, which lead to lower limits on the amount of non-thermal power. The total non-thermal energy is estimated by integrating the power in non-thermal electrons over the non-thermal emission duration, and may also be considered lower limits. For a typical microflare the non-thermal emission duration covers tens of seconds, while large flares may produce non-thermal energy over several minutes. The values of non-thermal energy are obtained by integrating the estimated power in non-thermal electrons over the 60 s fitting duration.

We find that the non-thermal power estimates range from a minimum of $\approx$ 1.2 $\times$ 10$^{26}$ erg s$^{-1}$ for event \#10 to a maximum of $\approx$ 1.35 $\times$ 10$^{27}$ erg s$^{-1}$ for event \#2. The exception is event \#5 which shows negligible non-thermal power when fitting the multi-thermal model. These estimated powers can be compared with the results of \citet{hannah2008}, who estimated the power in non-thermal electrons for 4236 microflare events under an assumption of isothermal plasma, finding a mean power of $P(>E_c) \approx 10^{26}$ erg s$^{-1}$ with a 2$\sigma$ range of 10$^{25}$ - 10$^{28}$ erg s$^{-1}$. 

Multiplying our estimated non-thermal powers by the RHESSI integration time, we find the estimated non-thermal energy content for these microflares is in the range 7 $\times$ 10$^{27}$ - 8 $\times$ 10$^{28}$ erg, again excluding event \#5 where the estimated non-thermal power is negligible. The values for each event are listed in Table \ref{energy_table}.

\subsection{Energy balance and comparison with an isothermal assumption}
\label{ebalance}

\begin{figure*}
\begin{center}
\includegraphics[width=8.5cm]{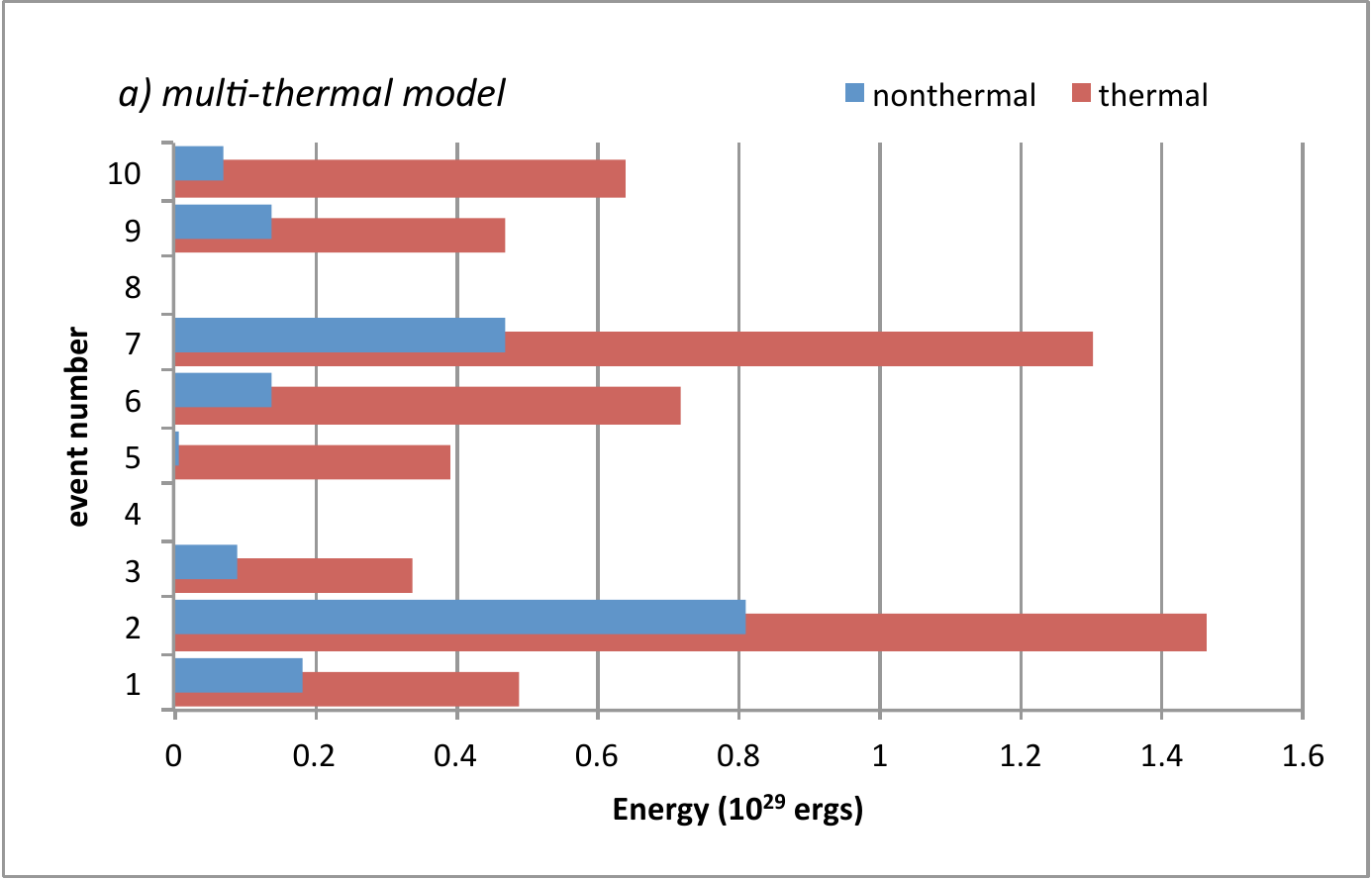}
\includegraphics[width=8.5cm]{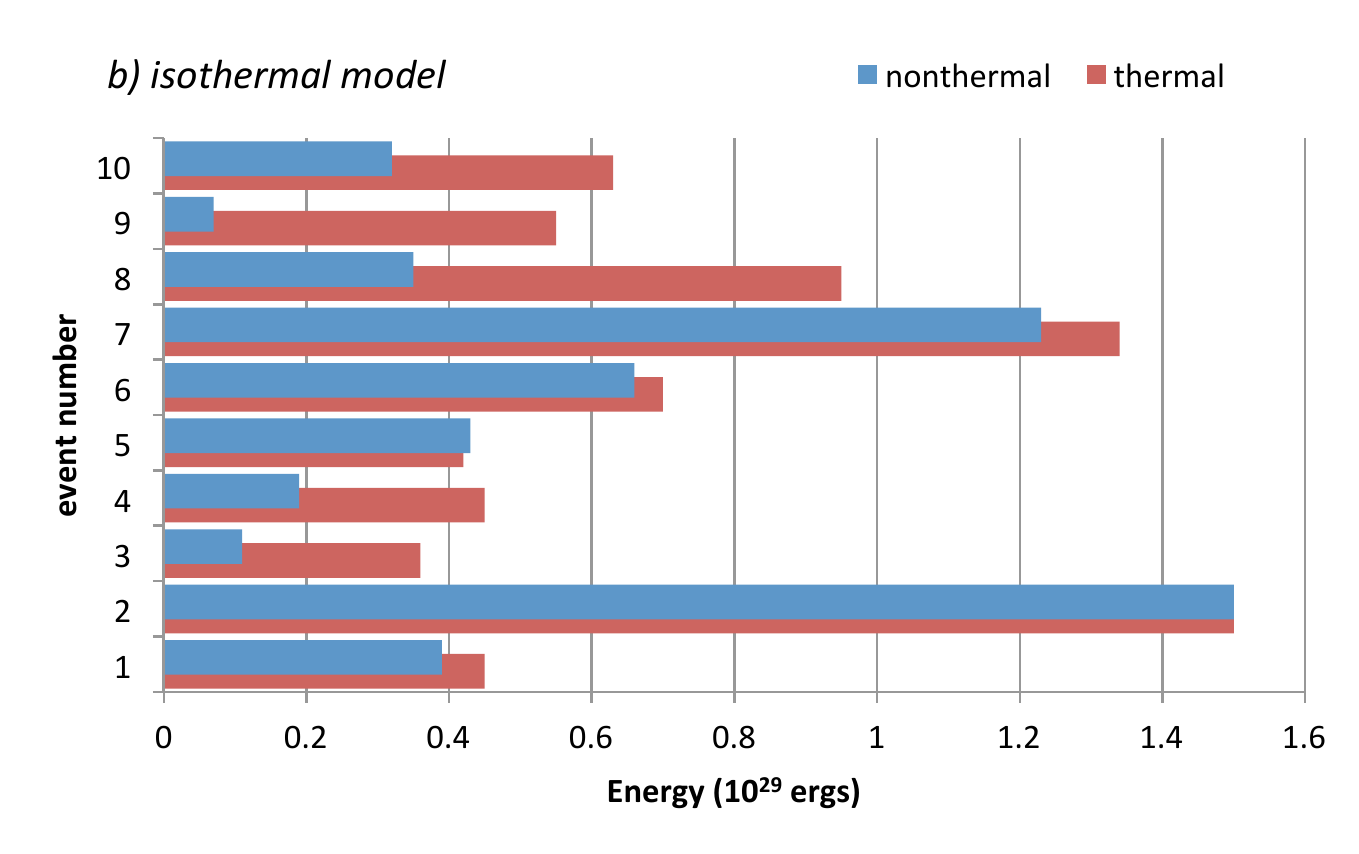}
\caption{Thermal and non-thermal energy estimates for each microflare under a) a multi-thermal assumption and b) an isothermal assumption.}
\label{energy_balance}
\end{center}
\end{figure*}

For comparison, the thermal and non-thermal energy content for these microflares was also estimated following the approach of \citet{hannah2008}, using an isothermal assumption with the addition of a thick target brehmsstrahlung component to model non-thermal particle emission. These results are shown in the right-most columns of Table \ref{energy_table}. 

\begin{figure}
\begin{center}
\includegraphics[width=8cm]{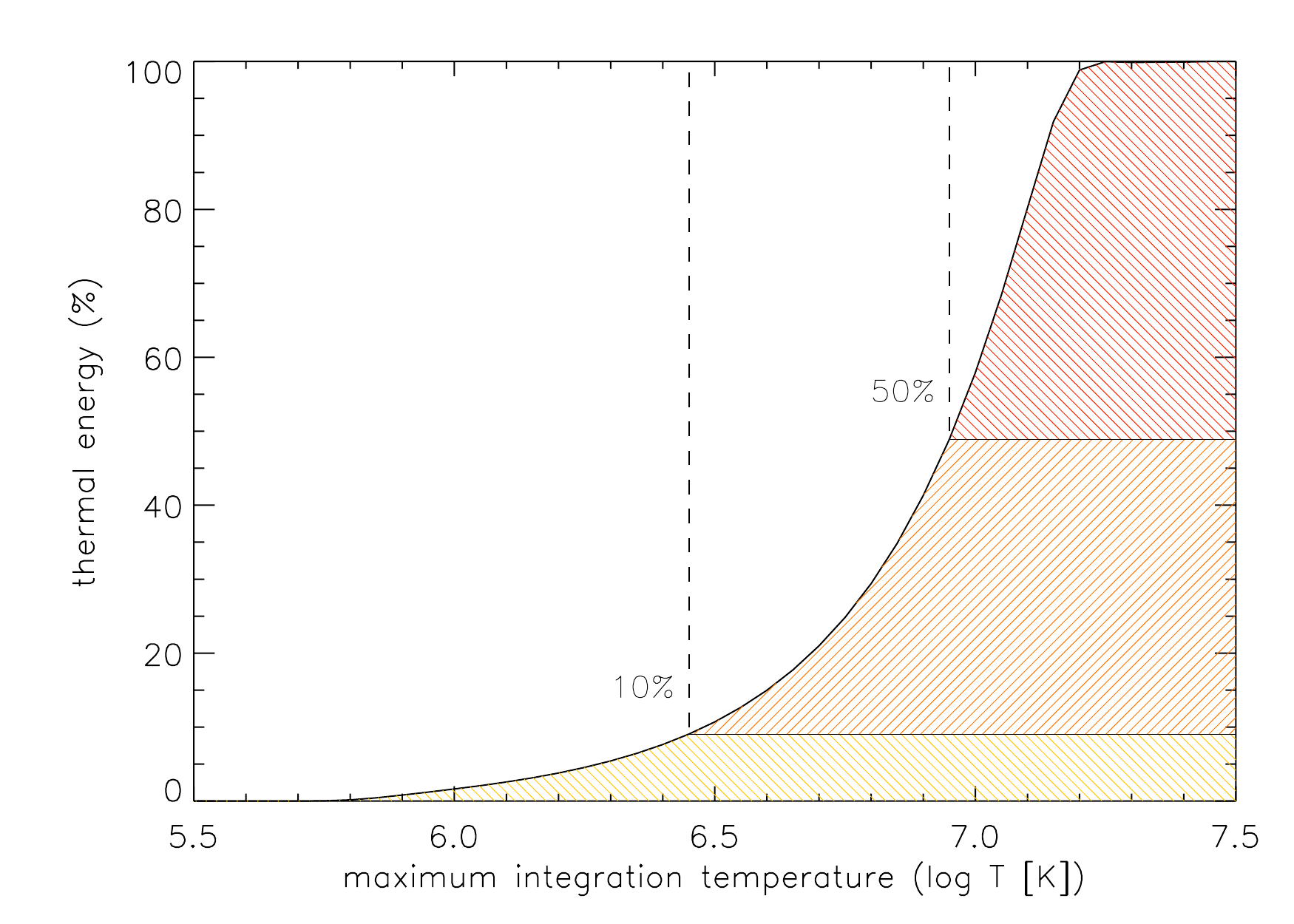}
\caption{Contribution to the total instantaneous thermal energy as a function of maximum integration temperature for a multi-thermal DEM given by Equation \ref{epstein_eqn}. In this example, the best-fit DEM from the 2011 June 5 microflare (\#1) was used. The curve in the figure is obtained by integrating over this best-fit DEM from low $T$, gradually increasing the upper limit of the integral, and comparing the obtained energies to the total energy obtained by integrating over all $T$.}
\label{thermal_accumulation}
\end{center}
\end{figure}

Here, we find that the estimated thermal energy content remains comparable with the estimates obtained under the multi-thermal model (see Figure \ref{energy_balance}). The similarity in thermal energy estimates may be explained by the distribution of contributions to the total instantaneous thermal energy in the multi-thermal case. Despite the multi-thermal nature of the plasma, most of the thermal energy is due to high-temperature plasma constrained by RHESSI observations. To illustrate this, in Figure \ref{thermal_accumulation} we integrate over the best-fit DEM obtained for microflare \#1, gradually increasing the upper temperature limit of the integral. Here, the emission below $\log T \approx 6.5$ contributes only 10\% of the total thermal energy estimate for this event, while almost 50\% of the total energy comes from emission above $\log T \approx 7.0$. 

By contrast, the amount of non-thermal energy is substantially increased in an isothermal scenario compared with a multi-thermal model (Figure \ref{energy_balance}). The estimated power in non-thermal electrons in this model varies between 1.2 $\times$ 10$^{26}$ erg s$^{-1}$ (event \#9) and 3.2 $\times$ 10$^{27}$ erg s$^{-1}$ (event \#4). Although above the mean power estimated in \citet{hannah2008}, these values are within their established 2$\sigma$ range.

For both the isothermal and the multi-thermal fits we compute the fraction $U_{nth} / U_{th}$. Figure \ref{energy_balance} illustrates the change in energy balance that results under a multi-thermal assumption; the minimum non-thermal energy is in general approximately 30\% of the estimated thermal energy. By contrast, under a traditional isothermal assumption the minimum estimated non-thermal energy is usually comparable with the thermal energy, the exception being event \#9 where $U_{nth} / U_{th} \approx 0.13$. This change is due to the multi-thermal spectrum pushing the thermal component and thus the low-energy cutoff $E_c$ to higher energies (see Table \ref{energy_table}). Here, the change in $E_c$ can be up to several keV, strongly affecting the estimated non-thermal energy contribution. It should be emphasised however that these are only lower limits on the level of non-thermal energy; the maximum cannot be constrained due to the uncertainty in $E_c$.

\section{Discussion and Conclusions}

In this paper we have, for the first time, combined EUV data from AIA and X-ray spectra from RHESSI to carry out joint forward-fitting of the differential emission measure distribution for a selection of microflares. Although the sample size of events here is small, the results presented here are strong and have implications for future studies of flare differential emission measure distributions and energetics. A study with a larger number of events is planned. We summarize our results below.

Firstly, we have found that a single Gaussian DEM distribution, although effective on AIA data in isolation, is unable to jointly fit AIA fluxes and RHESSI spectra for any of the studied microflares. This is due to the fact that such a function cannot decrease sufficiently steeply at high temperatures as required by the RHESSI X-ray observations while simultaneously remaining broad enough to accurately reproduce AIA observed fluxes at low temperatures. We therefore conclude that a single Gaussian DEM distribution should not be used in microflares.

Based on this finding we tested the hypothesis that a uniform DEM with a high temperature cut-off is an adequate distribution to describe the emission measure distribution of microflares. This hypothesis was tested using an Epstein profile with steepness parameter $n$ = 10 (see Equation \ref{epstein_eqn}). It was found that, for 80\% of the studied events, this simple model was able to fit both the AIA and RHESSI observations. Furthermore, for one event, $\#5$, the combined EUV and X-ray emission could be fit using only a multi-thermal plasma, suggesting a very weak or non-existent non-thermal component (see Figure \ref{epstein_subfig}). 

We conclude that for most cases, $80\%$, a simple boxcar distribution is a sufficient first-order approximation for describing microflare DEM within the uncertainties of the AIA and RHESSI instruments. However, for two of the studied events, a joint fit to the EUV and X-ray data was not possible using either the Gaussian or Epstein DEM models.  Consequently, the use of more complex DEM functions, such as a multiply peaked distribution, is justified in these cases.

The thermal energy and the total radiated energy of the microflares was investigated. It was found that, when a filling factor of $f \approx 1$ was assumed, the thermal energy exceeded the total radiated energy by two orders of magnitude. This was corroborated by estimating the radiated energy using background-subtracted GOES X-ray data for each event \citep{2005SoPh..227..231W, 2012ApJ...759...71E}. We can postulate that either conductive losses are dominating the cooling of the coronal plasma for these events, or the filling factor is significantly less than unity. By neglecting conductive losses and postulating that $L_{rad} \approx U_{th}$ we can estimate lower limits on the filling factor for these events to be $f \approx 5\times10^{-3} $. These estimates are plausible in view of the uncertainty surrounding the filling factor of the plasma \citep[e.g.][]{1997ApJ...478..799C}, although in reality some contribution to energy losses from conduction is expected during microflares, particularly during the rise phase of these events. Such low filling factors also appear to be inconsistent with more recent estimates of $f$ provided by \citet{2012ApJ...755...32G}, who suggested that $f$ should be of order 0.1 or 1. However, \citet{2012ApJ...755...32G} studied only large events. Typical filling factors for microflares may be different from those for large flares \citep[e.g.][]{1999ApJ...526..505M, 2011ApJ...736...75B}.

The non-thermal energy was also investigated. In Section \ref{energy} it was shown that the application of a multi-thermal model with a thick-target non-thermal component substantially reduces the lower limit on the level of energy in non-thermal particles during microflares (Figure \ref{energy_balance}) compared to assumption an isothermal component. In general it was found that, the minimum non-thermal energy content required was approximately 30\% of the estimated thermal energy. This is in contrast to the traditional isothermal model, where the thermal and non-thermal energies were found to be comparable. With this model therefore there is much more thermal energy compared to non-thermal energy though it must be noted that the upper limit on the non-thermal energy remains unconstrained since the low energy cutoff is not known. This suggests, though, that the standard model whereby the thermal emission in flares observed is due to chromospheric evaporation caused by the energy deposition of accelerated particles may not be appropriate in microflares. Instead, another source of energy, such as direct heating, accelerated protons, plasma waves, or DC-electric fields, may be the cause of the observed thermal energy.

\begin{acknowledgements}
 The authors are grateful to Dr Markus Aschwanden for helpful discussions and for making his software available within SSW/IDL, which provided the foundations of this work. We also thank Prof. A. G. Emslie for helpful discussions regarding flare energetics.
\end{acknowledgements}

\bibliographystyle{apj}
\bibliography{dem}

\end{document}